\documentclass[11pt]{article} 

\usepackage[utf8]{inputenc} 

\usepackage{cite}

\usepackage{hyperref}
\hypersetup{colorlinks=false}

\usepackage{geometry,amsmath} 
\usepackage{graphicx} 

\numberwithin{equation}{section}

\usepackage{booktabs} 
\usepackage{array} 
\usepackage{paralist} 
\usepackage{verbatim} 
\usepackage{subfig} 

\usepackage{float}

\usepackage{todonotes}

\usepackage{amssymb}
\usepackage{amsmath}
\usepackage{amsthm}
\usepackage{mathrsfs}

\theoremstyle{plain}
\newtheorem{thm}{Theorem}[section]

\newtheorem{exm}[thm]{Example}

\newtheorem{rem}[thm]{Remark}

\usepackage{fancyhdr} 
\usepackage{sectsty}
\allsectionsfont{\sffamily\mdseries\upshape} 

\title{Statistical Einstein manifolds of exponential families with group-invariant potential functions}

\date{} 
\author{Linyu Peng{$^{1,2,3}$}\footnote{Email: L.Peng@aoni.waseda.jp}
~  and Zhenning Zhang{$^{4}$}\footnote{Email: 
Zhangzhenning@bjut.edu.cn}\vspace{0.4cm}
\\
{\it 1. Waseda Institute for Advanced Study, Waseda University, Tokyo 169-8050, Japan}\\ 
{\it 2. Department of Applied Mechanics and Aerospace Engineering,} \\{\it Waseda University, Tokyo 169-8555, Japan}\\
{\it 3. School of Mathematics and Statistics, Beijing Institute of Technology, Beijing 100081, China}\\
{\it 4. Department of Mathematics, Beijing University of Technology, Beijing 100124, China
}
}

\begin{document}
\maketitle
\vspace{-0.4cm}
\begin{flushright}
{\it \Large Dedicated to Huafei Sun on the occasion of his sixtieth birthday.}
\end{flushright}
\vspace{0.2cm}

\begin{abstract}
This paper mainly contributes to a classification of statistical Einstein manifolds, namely statistical manifolds at the same time are Einstein manifolds. A statistical manifold is a Riemannian manifold, each of whose points is a probability distribution. With the Fisher information metric as a Riemannian metric, information geometry was developed to understand the intrinsic properties of statistical models, which play important roles in statistical inference, etc. Among all these models, exponential families is one of the most important kinds, whose geometric structures are fully determined by their potential functions. To classify  statistical Einstein manifolds, we derive partial differential equations for potential functions of exponential families; special solutions of these equations are obtained through the ansatz method as well as group-invariant solutions via reductions using Lie point symmetries.\\
{\bf Keywords:} information geometry; Einstein manifold; symmetry reduction; group-invariant solutions

\end{abstract}

\section{Introduction}
Information geometry was founded based on the applications of
differential geometry and Riemannian geometry into probability and statistics. 
The astonishing geometric structures are constructed based on the
Fisher information metric, which is viewed as a Riemannian metric
\cite{Chentsov:1982aa,Efron:1975aa,Rao:1945aa}. Decades later,
Amari and his collaborators \cite{Amari:1985aa,Amari:2000aa}
developed the dual geometric structures,
which are considered as the core of parametrised information geometry;
for more, see \cite{Arwini:2008aa,Murray:1993aa,Sun:2011aa,Sun:2016}. Beside
statistics and entropic dynamical models
\cite{Amari:1982ab,Amari:2011aa,Cafaro:2007aa,Cao:2008aa,Li:2015,Li:2008aa,Naudts:2009aa,Ohara:2010aa,Peng:2011aa,Peng:2007aa,Peng:2012sx},
information geometry has been successfully applied into other fields
as well, such as neural networks
\cite{Amari:1992aa,Amari:1995aa}, decoding \cite{Ikeda:2004aa}, control systems
\cite{Amari:1987aa,Zhang:2013,Zhang:2009aa,Zhong:2008aa}.

Meanwhile, there are scholars interested at the global properties of
statistical manifolds themselves, rather than the applications of
the theory of information geometry
\cite{Furuhata:2009aa,Kurose:2002aa,Takano:2006aa}. In this paper,
we investigate  statistical manifolds which are
Einstein manifolds at the same time, that is, the following equation holds
\cite{Besse:1987aa}
\begin{equation}\label{eq}
Ric=-\lambda g,
\end{equation}
where $\lambda$ is a constant, $Ric$ is the Ricci curvature tensor
and $g$ is the Riemannian or pseudo-Riemannian metric.
We call such statistical manifolds statistical Einstein manifolds.
In
information geometry, $g$ is the Fisher information 
metric, which is also called the Fisher information matrix.
 Exponential families, one of the most important type of statistical manifolds, are those whose probability density functions can be written in terms of the so-called potential functions; see Section 3 for an accurate definition. A well-known example of exponential family is the normal distribution, also called Gaussian distribution. Geometric quantities of exponential families, e.g. Fisher information metric, curvatures, can be fully determined by the potential functions. Eq. \eqref{eq} for the metric $g$ then becomes a differential equation of potential functions. 

The complicity makes it extremely challenging to solve Eq. \eqref{eq} or the equation for potential functions in general. In the current paper, we try to obtain special solutions in particular via reductions using Lie point symmetries; see for instance \cite{Hydon:2000aa,Olver:1993aa,Bluman:1974,Bluman:2010,Bluman:1989}. First introduced by Lie during 1880s, symmetry method has been greatly exploited and applied in various aspects. Continuous symmetries, in plain words, are local transformations unchanging the shape of an object; for differential equations, the object is the set of solutions. It is believed that every solvable differential equation has symmetries behind it. 

The paper is organised as follows. In Section 2, we briefly review the theory of Lie point symmetries and group-invariant solutions. We will introduce the main concepts of information geometry in Section 3 and construct statistical Einstein manifolds in Section 4; some examples are provided. In Section 5, we obtain some two dimensional statistical Einstein manifolds by solving the corresponding equation \eqref{eq} using an ansatz method and the symmetry method. The last section contributes to conclusions and open questions.

\section{Lie point symmetries and group-invariant solutions of partial differential equations}
%
%
%
This section is mostly based on \cite{Olver:1993aa}; see also \cite{Hydon:2000aa,Bluman:1974,Bluman:2010,Bluman:1989}. In general, let $x=(x^1,x^2,\ldots,x^n)$ be the $n$ independent variables and let $u=(u^1,u^2,\ldots,u^m)$ be the $m$ dependent variables. In the theory of information geometry below, they will be replaced by the parameters $\theta$ and potential function $\psi(\theta)$, respectively. Consider the following system of differential equations
\begin{equation}\label{des}
F_{k}(x,u,u^{\alpha}_i,\ldots)=0, ~~ k=1,2,\ldots, K,
\end{equation}
where each $F_k$ is a function of finitely-many arguments.
Here $u_i^{\alpha}:=\frac{\partial u^{\alpha}}{\partial x^i}$ denotes first order partial derivatives and we use the shorthand notation $u_{\bold{J}}^{\alpha}$ to denote partial derivatives:
\begin{equation*}
u_{\bold{J}}^{\alpha}:=\frac{\partial^{|{\bold{J}}|} u^{\alpha}}{\partial (x^1)^{j_1}\partial (x^2)^{j_2}\cdots \partial (x^n)^{j_n}}
\end{equation*}
for ${\bold{J}}=(j_1,j_2,\ldots, j_n)$ with non-negative entries and $|{\bold{J}}|=j_1+j_2+\cdots+j_n$.
 For a one-parameter Lie group $G$ of transformations: 
\begin{equation}
\begin{aligned}
&\widetilde{x}=\widetilde{x}(\varepsilon,x,u),~\widetilde{u}=\widetilde{u}(\varepsilon,x,u);\\
&\widetilde{x}|_{\varepsilon=e}=x,~\widetilde{u}|_{\varepsilon=e}=u,
\end{aligned}
\end{equation}
the associated infinitesimal generator is
\begin{equation}
X=\xi^i(x,u)\frac{\partial}{\partial x^i}+\eta^{\alpha}(x,u)\frac{\partial}{\partial u^{\alpha}},
\end{equation}
where
\begin{equation}
\xi^i:=\frac{\operatorname{d}}{\operatorname{d}\!\varepsilon}\Big|_{\varepsilon=e}\widetilde{x}^i,~\eta^{\alpha}=\frac{\operatorname{d}}{\operatorname{d}\!\varepsilon}\Big|_{\varepsilon=e}\widetilde{u}^{\alpha}.
\end{equation}
Here $e$ is the identity element of $G$. 
The Einstein
summation convention is used here and all through the paper.  This group action can be prolonged to partial derivatives of $u$, leading to the prolongation of the vector field $X$\cite{Olver:1993aa}:
\begin{equation}
\bold{pr}X=\xi^iD_i+\cdots+D_{\bold{J}}(\eta^{\alpha}-u^{\alpha}_i\xi^i)\frac{\partial}{\partial u_{\bold{J}}^{\alpha}}+\cdots.
\end{equation}
Here $Q^{\alpha}:=\eta^{\alpha}-u^{\alpha}_i\xi^i$ are called characteristics of the transformations; $D_i$ is the total derivative with respect to $x^i$. The multi-index notation $D_{\bold{J}}$ is a composition of total derivatives:
\begin{equation*}
D_{\bold{J}}=D_{x^1}^{j_1}D_{x^2}^{j_2}\cdots D_{x^n}^{j_n}.
\end{equation*} 
This group of transformations is a symmetry group for the system of differential equations (\ref{des}) if and only if the linearized symmetry condition (LSC) is satisfied 
\begin{equation}\label{lsc}
\bold{pr}X(F_k)=0~~\text{on all solutions of  (\ref{des}).}
\end{equation}
Such groups of symmetries with $\xi$ and $\eta$ being independent from partial derivatives are called groups of Lie point symmetries. Otherwise, they are called groups of higher symmetries.

\begin{exm}[Heat equation]
Consider the heat equation
\begin{equation}
u_t=u_{xx}.
\end{equation}
Let $X=\xi^t(t,x,u)\partial_t+\xi^x(t,x,u)\partial_x+\eta(t,x,u)\partial_u$ where $\partial_t=\frac{\partial}{\partial t}$, $\partial_x=\frac{\partial}{\partial x}$ and so forth; its prolongation is
\begin{equation}
\bold{pr}X=\xi^t\partial_t+\xi^x\partial_x+\eta\partial_u+\eta^t\partial_{u_t}+\eta^{xx}\partial_{u_{xx}}+\cdots,
\end{equation}
where 
\begin{equation}
\eta^t=u_{tt}\xi^t+u_{tx}\xi^x+D_tQ,~\eta^{xx}=u_{xxx}\xi^x+u_{txx}\xi^t+D_x^2Q.
\end{equation}
Here the characteristic is $Q=\eta-\xi^tu_t-\xi^xu_x$. Then the LSC, that is $\bold{pr}X(u_t-u_{xx})=0$ when $u_t-u_{xx}=0$, gives
\begin{equation}
\eta^t-\eta^{xx}=0~\text{when}~u_t-u_{xx}=0.
\end{equation}
Substitute $u_t=u_{xx}$ (as well as $u_{tt}=u_{xxxx}$, $u_{tx}=u_{xxx}$ and so on) to the equality $\eta^t-\eta^{xx}=0$.We will get a polynomial with respect to partial derivatives $u_x$, $u_{xx}$ and so on. All partial derivatives along the $t$ direction will be replaced by partial derivatives about $x$. By equating all independent coefficients of such a polynomial to zero, we get a system of differential equations about $\xi^t$, $\xi^x$ and $\eta$. Solve them we will obtain all infinitesimal generators. It is spanned by the six vector fields
\begin{equation*}
\begin{aligned}
&X_1=\partial_x,~~X_2=\partial_t,~~X_3=u\partial_u,~~X_4=x\partial_x+2t\partial_t,\\
&X_5=2t\partial_x-xu\partial_u,~~X_6=4tx\partial_x+4t^2\partial_t-(x^2+2t)u\partial_u,
\end{aligned}
\end{equation*}
and the infinite-dimensional subalgebra
\begin{equation*}
h(t,x)\partial_u,
\end{equation*}
where $h(t,x)$ is an arbitrary solution of the heat equation.

\end{exm}

 Note that computer algebra packages are available for computing symmetries of differential equations using Maple, e.g. DESOLVII \cite{VuJeCa:2012} and GeM \cite{Cheviakov:2007}.

Having the infinitesimal generators in hand, we are able to construct differential invariants and hence group-invariant solutions. A function $f(x,u)$ is called invariant with respect to a vector field $X$ if for any section $(x,u(x))$, the equality holds identically
\begin{equation}
X(f)=0.
\end{equation}
Expanding this equality leads to a linear partial differential equation about $f$,
\begin{equation}
\eta^{\alpha}\frac{\partial f}{\partial u^{\alpha}}+\xi^i\frac{\partial f}{\partial x^i}=0,
\end{equation}
which can often be solved using the method of characteristics, namely by solving
\begin{equation}\label{det}
\frac{\operatorname{d}\!x^1}{\xi^1}=\cdots=\frac{\operatorname{d}\!x^n}{\xi^n}=\frac{\operatorname{d}\!u^1}{\eta^1}=\cdots=\frac{\operatorname{d}\!u^m}{\eta^m},
\end{equation}
if all coefficients $\xi$ and $\eta$ are nonzero.
Solutions of \eqref{det} will provide us differential invariants, e.g. $h^1,h^2,\ldots$ Any invariant function $f(x,u)$ is hence rewritten as $f(h^1,h^2,\ldots)$. Consider such invariants as new coordinates for the original system (\ref{des}) and we obtain a reduced system. In specific cases, the reduced system can be solved exactly. Multi-reduction is also possible; see, for instance \cite{Hydon:2000aa,Olver:1993aa}. Let us find out how it works in practice through the following example.

\begin{exm}[Heat equation cont.]
For instance, consider the  combination 
\begin{equation}
X=X_4+aX_3=x\partial_x+2t\partial_t+au\partial_u,
\end{equation}
where $a$ is a nonzero constant.
The equations for determining differential invariants \eqref{det} become 
\begin{equation}
\frac{\operatorname{d}\!x}{x}=\frac{\operatorname{d}\! t}{2t}=\frac{\operatorname{d}\!u}{au}.
\end{equation}
Those equations can be solved exactly. For example consider the first equality and we get $x=C\sqrt{t}$, which implies (by solving for the constant) that 
\begin{equation}
y:=\frac{x}{\sqrt{t}}
\end{equation}
is a differential invariant. Similarly we have another differential invariant
\begin{equation}
v=t^{-a}u.
\end{equation}
The variable $v$ is considered as a function of $y$ only, namely $v=v(y)$. Therefore, we can compute 
\begin{equation}
u_t=D_t(v(y)t^a)=v'y_tt^a+vat^{a-1}
\end{equation}
and 
\begin{equation}
u_{xx}=D_x^2(v(y)t^a)=t^aD_x^2v(y)=t^aD_x(v'y_x)=t^a(v''y_x^2+v'y_{xx}).
\end{equation}
Substitute these back to the heat equation $u_t=u_{xx}$ and we obtain an ordinary differential equation 
\begin{equation}
v''+\frac{1}{2}yv'-av=0.
\end{equation}
Its general solution is
\begin{equation}
    v(y)=y\exp\left(-\frac{y^2}{4}\right)\left\{c_1M\left(a+1,\frac{3}{2},\frac{y^2}{4}\right)+c_2U\left(a+1,\frac{3}{2},\frac{y^2}{4}\right)\right\}.
\end{equation}
where $M$ and $U$ represent the Kummer's and Tricomi's confluent hypergeometric functions, respectively, e.g. \cite{AS1964}, and $c_1$ and $c_2$ are integration constants.
 Changing them back to the original coordinates, we get an invariant solution for the heat equation
\begin{equation*}
u(x,t)=xt^{a-\frac{1}{2}}\exp\left(-\frac{x^2}{4t}\right)\left\{c_1M\left(a+1,\frac{3}{2},\frac{x^2}{4t}\right)+c_2U\left(a+1,\frac{3}{2},\frac{x^2}{4t}\right)\right\}.
\end{equation*}
\end{exm}

In the next section, we will derive a system of differential equations for the potential functions of exponential families of probability density functions and investigate its group-invariant solutions. They contribute to statistical Einstein manifolds in information geometry.

\section{Information geometry and exponential families}
Although geometric analysis of any general finite parametric
distribution can be developed, here we only consider  regular
distributions, whose probability density functions (pdfs)
$p(x;\theta)$ with parameters $\theta=(\theta^1,\theta^2,\ldots,\theta^n)\in\Theta$ satisfy the following regularity conditions:
\begin{enumerate}
\item $\Theta$ is a subset of $\mathbb{R}^n$ such that for each $x$,
the mapping $\theta\mapsto p(x;\theta)$ is smooth.
\item The order of integration and differentiation can be freely
rearranged. For instance,
\begin{equation}
\int \partial_ip(x;\theta)\operatorname{d}\!x=
\partial_i\int p(x;\theta)\operatorname{d}\!x=\partial_i 1=0,
\end{equation}
where $\partial_i=\frac{\partial}{\partial \theta^i}$. For
discrete distributions, we simply replace the integration by summation.
\item Different parameters stand for different pdfs, that is,
$\theta_1\neq\theta_2$ implies that $p(x;\theta_1)$ and
$p(x;\theta_2)$ are different pdfs. Moreover, every parameter
$\theta$ possesses a common support set where $p(x;\theta)> 0$.
\end{enumerate}

The family $S$ of distribution represented by the pdf $p(x;\theta)$
is called an $n$-dimensional statistical model
\begin{equation}
S=\left\{p(x;\theta)\mid \theta\in\Theta\subset \mathbb{R}^n\right\}.
\end{equation}
There are many examples of statistical models, such as Poisson
distribution \cite{Amari:2000aa}, inverse Gamma distribution
\cite{Li:2008aa} and Pareto distribution \cite{Peng:2007aa}. The
following examples are particularly interesting to us as both of them correspond to statistical Einstein manifolds.
\begin{exm}[Normal distribution or Gaussian distribution]
\label{nd} The dimension is $n=2$, $x\in\mathbb{R}$, and the pdf is
\begin{equation}
p(x;\theta)=\frac{1}{\sqrt{2\pi}\sigma}\exp\left\{-\frac{(x-\mu)^2}{2\sigma^2}\right\},
\end{equation}
where the parameter space is $\Theta=\{\theta=(\mu,\sigma)\mid\mu\in \mathbb{R},\sigma>0\}$. The parameters $\mu$ and $\sigma$ are the mean and standard deviation.
\end{exm}

\begin{exm}[Weibull distribution]
\label{wd} The dimension $n=2$, $x\geq0$, and the pdf is given by
\begin{equation}
p(x;\theta)=\frac{b}{a}\left(\frac{x}{a}\right)^{b-1}\exp\left\{-\left(\frac{x}{a}\right)^b\right\},
\end{equation}
where $\Theta=\{\theta=(a,b)\mid a>0,b>0\}$. Here $b$ is the shape parameter and $a$ is the scale parameter of the distribution.
\end{exm}

\subsection{Fisher information metric and dual geometric structures}
The Fisher information metric of a statistical model $S$ is defined as
\begin{equation}
g_{ij}(\theta):=E[\partial_i l_{\theta}~\partial_j l_{\theta}]=\int
\partial_i l(x;\theta)\partial_j l(x;\theta)p(x;\theta)\operatorname{d}\!x,
\end{equation}
where $l_{\theta}=l(x;\theta)=\ln p(x;\theta)$ and $E$ means the
expectation. For simplicity, we sometimes write $g_{ij}$
instead of $g_{ij}(\theta)$ and so forth. The positivity of $(g_{ij})$ makes $S$ to be
a Riemannian manifold; see e.g. \cite{Amari:2000aa,Sun:2011aa,Sun:2016}. There exists a one-parameter group of
dual connections $\nabla^{(\alpha)}$, the so called $\alpha$-connection,
such that $\nabla^{(0)}$ is the unique Riemannian connection
corresponding to $g$. The duality is illustrated by 
\begin{equation*}
Xg(Y,Z)=g\left(\nabla^{(\alpha)}_XY,Z\right)+g\left(Y,\nabla^{(-\alpha)}_XZ\right),
\end{equation*}
where $X,Y,Z$ are vector fields on the manifold.
The coefficients of $\alpha$-connection
are given by
\begin{equation}
\begin{aligned}
\Gamma_{ij,k}^{(\alpha)}:&=E[\partial_i\partial_j
l_{\theta}~\partial_k l_{\theta}]+\frac{1-\alpha}{2}E[\partial_i
l_{\theta}~\partial_j l_{\theta}~\partial_k l_{\theta}]\\
&=\Gamma_{ij,k}^{(0)}+\frac{\alpha}{2}T_{ijk},
\end{aligned}
\end{equation}
with $\Gamma_{ij,k}^{(0)}$ the coefficients of Riemannian connection
$\nabla^{(0)}$ and
\begin{equation}
T_{ijk}:=E[\partial_i l_{\theta}~\partial_j l_{\theta}~\partial_k
l_{\theta}].
\end{equation}
Similarly to the Riemannian case, we have the
$\alpha$-curvature tensor in components \cite{Carmo:1992aa,Petersen:2006aa}
\begin{equation} 
R^{(\alpha)l}_{kij}=\partial_i\Gamma^{(\alpha)l}_{kj}-\partial_j\Gamma^{(\alpha)l}_{ki}
+\Gamma^{(\alpha)h}_{kj}\Gamma^{(\alpha)l}_{hi}-\Gamma^{(\alpha)h}_{ki}\Gamma^{(\alpha)l}_{hj},
\end{equation}
which can also be written as
$R_{klij}^{(\alpha)}=R_{kij}^{(\alpha)s}g_{sl}$. The
$\alpha$-Ricci curvature tensor and $\alpha$-scalar curvature are
defined respectively as
\begin{equation}
Ric_{ij}^{(\alpha)}=R_{iklj}^{(\alpha)}g^{kl}
\end{equation}
and
\begin{equation}
K^{(\alpha)}=Ric_{ij}^{(\alpha)}g^{ij},
\end{equation}
where $(g^{ij})$ is the inverse of the Fisher information metric. The
$\alpha$-sectional curvature tensor is given by
\begin{equation}
\kappa_{ij}^{(\alpha)}=-\frac{R_{ijij}^{(\alpha)}}{g_{ii}g_{jj}-g_{ij}^2},\quad i\neq j.
\end{equation}

\subsection{Exponential families}
Pdfs of exponential families, one of the most important class of statistical models, can be written as
\begin{equation}
p(x;\theta)=\exp\left\{F_i(x)\theta^i+C(x)-\psi(\theta)\right\},
\end{equation}
which are assumed to satisfy the regularity conditions. The functions $F_i(x)$ are smooth with respect to $x$. From the normalization condition $\int
p(x;\theta)\operatorname{d}\!x=1$, the potential function is given by
\begin{equation}
\psi(\theta)=\ln\int\exp\left\{F_i(x)\theta^i+C(x)\right\}\operatorname{d}\!x.
\end{equation}
The normal distribution, e.g. Example \ref{nd}, is a well-known example of exponential families however the Weibull distribution, e.g. Example
\ref{wd}, does not belong to exponential families. Geometric quantities of exponential families can be neatly written using the potential functions
$\psi(\theta)$.
\begin{thm}[\cite{Amari:1982ab,Amari:2000aa,Sun:2016}]\label{thm:expfam}
For exponential families:
\begin{enumerate}
\item The Fisher information metric is given by $g_{ij}=\partial_i\partial_j
\psi(\theta)$, which implies that potential functions are convex.
\item The coefficients of $\alpha$-connection are
\begin{equation}
\Gamma_{ij,k}^{(\alpha)}=\frac{1-\alpha}{2}T_{ijk},
\end{equation}
where
\begin{equation}
T_{ijk}=\partial_i\partial_j\partial_k\psi(\theta).
\end{equation}
\item The components of $\alpha$-curvature tensor are
\begin{equation}
R_{ijkl}^{(\alpha)}=\frac{1-\alpha^2}{4}(T_{kmi}T_{jln}-T_{kmj}T_{iln})g^{mn}.
\end{equation}
\end{enumerate}
\end{thm}

\section{Statistical Einstein manifolds of exponential families}
An Einstein manifold is a Riemannian or pseudo-Riemannian manifold
which satisfies the equation \eqref{eq}, namely
\begin{equation*}
Ric=-\lambda g.
\end{equation*}
\begin{exm}[Cont.]\label{ex1c}
For Example \ref{nd}, it is an exponential family as we can rewrite
the pdf as
\begin{equation}
p(x;\theta)=\exp\left\{\frac{\mu}{\sigma^2}x-\frac{1}{2\sigma^2}x^2-\frac{\mu^2}{2\sigma^2}-\ln\sigma
-\frac{1}{2}\ln 2\pi\right\}.
\end{equation}
Introducing the so-called natural parameters
$(\theta^1=\frac{\mu}{\sigma^2},\theta^2=-\frac{1}{2\sigma^2})$,
and a $2$-tuple $(F_1(x)=x,F_2(x)=x^2)$, we have
\begin{equation}
p(x;\theta)=\exp\left\{F_i(x)\theta^i-\psi(\theta)\right\},
\end{equation}
where
\begin{equation}
\psi(\theta)=-\frac{(\theta^1)^2}{4\theta^2}-\frac{1}{2}\ln(-\theta^2)+\frac{1}{2}\ln\pi.
\end{equation}
Therefore, we can get the Fisher information metric and the Ricci curvature
tensor as
\begin{equation}
g=\left(\begin{array}{cc} \sigma^2&2\mu\sigma^2\\
2\mu\sigma^2&4\mu^2\sigma^2+2\sigma^4
\end{array}
\right)
\end{equation}
and
\begin{equation}
Ric=\left(\begin{array}{cc} -\frac{\sigma^2}{2}&-\mu\sigma^2\\
-\mu\sigma^2&-2\mu^2\sigma^2-\sigma^4
\end{array} \right).
\end{equation}
It is obvious that they satisfy the equation \eqref{eq}
\begin{equation}
Ric=-\frac{1}{2}g=\kappa_{12} g,
\end{equation}
with $-\lambda=\kappa_{12}=-\frac{1}{2}$ the sectional curvature.
\end{exm}

\begin{exm}[Cont.]\label{ex2c}
Example \ref{wd} does not belong to exponential families, however, it also corresponds to a statistical Einstein manifold. By viewing $\theta=(a,b)$ as local coordinates, the
Fisher information metric is \cite{Cao:2008aa}
\begin{equation}
g=\left(\begin{array}{cc} \frac{b^2}{a^2}&-\frac{1-\xi}{a}\\
-\frac{1-\xi}{a}&\frac{\xi^2-2\xi+\frac{\pi^2}{6}+1}{b^2}
\end{array}\right),
\end{equation}
where $\xi$ is the Euler's constant. The Ricci curvature is given by
\begin{equation}
Ric=\left(\begin{array}{cc} \frac{6b^2}{\pi^2a^2}&-\frac{6(1-\xi)}{\pi^2a}\\
-\frac{6(1-\xi)}{\pi^2a}&\frac{6(\xi^2-2\xi+\frac{\pi^2}{6}+1)}{\pi^2b^2}
\end{array}\right).
\end{equation}
Therefore,
\begin{equation}
Ric=\frac{6}{\pi^2}g=\kappa_{12} g,
\end{equation}
where $-\lambda=\kappa_{12}=\frac{6}{\pi^2}$ is the Gaussian curvature.
This fact implies that the Weibull distribution manifold is an Einstein manifold
as well.
\end{exm}

For exponential families, we are able to write down the equation \eqref{eq} explicitly for the potential functions:
\begin{equation}\label{eq:Ricg}
R^{(0)}_{iklj}g^{kl}=-\lambda g_{ij},
\end{equation}
that is,
\begin{equation}
\sum_{m,n,k,l}\frac{1}{4}\{\partial_i\partial_l\partial_m\psi(\theta)\partial_k\partial_j\partial_n\psi(\theta)
-\partial_k\partial_l\partial_m\psi(\theta)
\partial_i\partial_j\partial_n\psi(\theta)\}g^{mn}g^{kl}=-\lambda\partial_i\partial_j\psi(\theta).
\end{equation}
\begin{rem}
Note that for exponential families, equation \eqref{eq:Ricg} actually holds for all $\alpha$ if it holds for $\alpha=0$ as a consequence of Theorem \ref{thm:expfam}. Without loss of generality,  we will set $\alpha=0$.
\end{rem}

For dimension two, in particular, we have the Fisher information metric
\begin{equation}
g=\left(\begin{array}{cc}
\partial_1\partial_1\psi&\partial_1\partial_2\psi\\
\partial_2\partial_1\psi&\partial_2\partial_2\psi
\end{array}\right),
\end{equation}
and therefore
$\det(g)=\partial_1\partial_1\psi\partial_2\partial_2\psi-(\partial_1\partial_2\psi)^2$.
The Riemannian curvature tensor is given by 
\begin{equation}
\begin{aligned}\label{rct}
R^{(0)}_{1212}=&~\frac{1}{4}\sum_{m,n}(\partial_1\partial_1\partial_m\psi\partial_2\partial_2\partial_n\psi-\partial_1\partial_2\partial_m\psi
\partial_1\partial_2\partial_n\psi)g^{mn}\\
=&~\frac{1}{4\det(g)}\Big\{\partial_1\partial_1\psi[\partial_1\partial_1\partial_2\psi\partial_2\partial_2\partial_2\psi
-(\partial_1\partial_2\partial_2\psi)^2]\\
&~-\partial_1\partial_2\psi[\partial_1\partial_1\partial_1\psi\partial_2\partial_2\partial_2\psi
-\partial_1\partial_1\partial_2\psi\partial_1\partial_2\partial_2\psi]\\
&~+\partial_2\partial_2\psi[\partial_1\partial_1\partial_1\psi\partial_1\partial_2\partial_2\psi
-(\partial_1\partial_1\partial_2\psi)^2]\Big\}.
\end{aligned}
\end{equation}
The components of the Ricci curvature are 
\begin{equation}
Ric_{11}^{(0)}=-\frac{\partial_1\partial_1\psi
}{\det(g)}R^{(0)}_{1212},~~Ric_{21}^{(0)}=Ric_{12}^{(0)}=-\frac{\partial_1\partial_2\psi
}{\det(g)}R^{(0)}_{1212},~~Ric_{22}^{(0)}=-\frac{\partial_2\partial_2\psi
}{\det(g)}R^{(0)}_{1212}.
\end{equation}
Therefore, the equation has the following equivalent representation
\begin{equation}\label{eeq}
-\frac{ R_{1212}^{(0)}}{\det(g)}=-\lambda,
\end{equation}
the left hand side of that is the sectional curvature. So in this
case, we can replace the constant $-\lambda$ by the
curvature $\kappa_{12}$. In this situation, the Einstein manifold is equivalent to
a manifold with constant curvature.
\begin{thm}[\cite{Besse:1987aa}]
A 2 or 3-dimensional pseudo-Riemannian manifold is Einstein if and
only if it has constant sectional curvature.
\end{thm}
In the next section, some special solutions of the equation
(\ref{eeq}) are obtained, in particular group-invariant solutions via Lie point symmetries, that give us the potential functions for statistical Einstein manifolds of exponential families.


\section{Two dimensional statistical Einstein manifolds of exponential families}
In this section, we obtain some special solutions of the equation (\ref{eeq})
for two dimensional exponential families. There is a prerequisite
that the potential function is convex. For example, it is
impossible to find any traveling wave solution of the form
$\psi(\theta^1,\theta^2)=f(\theta^1-c\theta^2)$ with $c$ a constant,
since $\det(g)=0$ for any $c$ and any smooth function $f$.

\subsection{The ansatz method}
 First, we try the ansatz method, namely by pre-assuming the shapes of potential functions $\psi(\theta^1,\theta^2)$ and solving the relevant equation \eqref{eeq}. 
 
 {\bf 1. Summation of two arbitrary functions, i.e. $\psi(\theta^1,\theta^2)=f(\theta^1)+h(\theta^2)$.}\\
In this case, the Riemannian curvature tensor always vanishes by
(\ref{rct}), and the Fisher information metric is $g=\operatorname{diag}(f'',h'')$. Therefore, the parameter must be $\lambda=0$  and any functions $f$ and $h$ satisfying the convexity condition that $f''> 0$ and $h''>0$ for all parameters $\theta^1$ and $\theta^2$ provide a potential function for a statistical Einstein manifold. 

{\bf 2. Summation of an arbitrary function and a traveling wave
function, i.e.
$\psi(\theta^1,\theta^2)=f(\theta^1)+h(\theta^1-c\theta^2)$ with $c$
a nonzero constant.}\\
The
Riemannian curvature tensor vanishes as well, and the Fisher information metric
is given by
\begin{equation}
g=\left(\begin{array}{cc} f''+h''&-ch''\\
-ch''&c^2h''
\end{array}
\right).
\end{equation}
Therefore $\lambda=0$, and any set of functions $f$ and $h$ satisfying the
convexity condition $f''+(1+c^2)h''>0$ and $f''h''>0$ provides a solution. 

{\bf 3. Multiplication of two arbitrary functions, i.e. $\psi(\theta^1,\theta^2)=f(\theta^1)h(\theta^2)$.}\\
In this case, we are only able to obtain some particular solutions.
Now the Fisher information metric is
\begin{equation}
g=\left(
\begin{array}{cc}
f''h&f'h'\\
f'h'&fh''
\end{array}
\right),
\end{equation}
which implies the convexity condition that $f''h+fh''>0$ and
$\det(g)=fhf''h''-(f'h')^2>0$ for all $\theta^1$ and $\theta^2$ in some domain $\Theta\subset \mathbb{R}^2$. The equation (\ref{eeq})
becomes 
\begin{equation}
4\lambda\det(g)^2+f(f''^2-f'f''')hh'h'''+f'(ff'''-f'f'')hh''^2+f''(f'^2-ff'')h'^2h''=0.
\end{equation}

When $\lambda=0$, the equation above is reduced to the zero-curvature equation
\begin{equation}\label{case30}
f(f''^2-f'f''')hh'h'''+f'(ff'''-f'f'')hh''^2+f''(f'^2-ff'')h'^2h''=0,
\end{equation}
and it can be solved explicitly as follows. Define $A_1(\theta^1)=f(f''^2-f'f''')$, $A_2(\theta^1)=f'(ff'''-f'f'')$ and $A_3(\theta^1)=f''(f'^2-ff'')$. As $h''\neq0$, rewrite the equation above as
\begin{equation}\label{case301}
A_1(\theta^1)hh'h'''+A_2(\theta^1)hh''^2+A_3(\theta^1)h'^2h''=0,
\end{equation}
 and divide $hh''^2$ on both sides
\begin{equation}
A_1(\theta^1)\frac{h'h'''}{h''^2}+A_2(\theta^1)+A_3(\theta^1)\frac{h'^2}{hh''}=0.
\end{equation}
Then we differentiate it with respect to $\theta^2$ to get
\begin{equation}
A_1(\theta^1)\left(\frac{h'h'''}{h''^2}\right)'+A_3(\theta^1)\left(\frac{h'^2}{hh''}\right)'=0.
\end{equation}
Therefore, noting that $h'\neq0$ from the convexity condition, 
\begin{itemize}
\item If $\frac{h'^2}{hh''}=c_1\neq 0$, that is $\left(\frac{h'^2}{hh''}\right)'=0$, it can be solved as follows:
\begin{itemize}
\item When $c_1=1$, we have
 \begin{equation}
h(\theta^2)=c_3\exp(c_2\theta^2)
\end{equation}
 and hence $\left(\frac{h'h'''}{h''^2}\right)'=0$. The
equation (\ref{case30}) holds constantly for any function $f$. Therefore, a solution would be
\begin{equation}
\psi(\theta^1,\theta^2)=f(\theta^1)\exp(c_2\theta^2),
\end{equation}
where $f$ is an arbitrary function such that the convexity condition is satisfied.
\item Otherwise when $c_1\neq 1$, we have
\begin{equation}
h(\theta^2)=
\left(c_2\theta^2-c_3\right)^{\frac{c_1}{c_1-1}}
\end{equation}
and again $\left(\frac{h'h'''}{h''^2}\right)'=0$ holds for the solution obtained above. Equation (\ref{case301}) gives an extra equation of $f$
\begin{equation*}
(2-c_1)A_1(\theta^1)+A_2(\theta^1)+c_1A_3(\theta^1)=0,
\end{equation*}
whose general solution is
\begin{equation}
f(\theta^1)=c_6(\theta^1-c_5)^{c_4}.
\end{equation}
In general, the potential function is not convex. However, it is possible to find special convex solutions, e.g.
\begin{equation}
\psi(\theta^1,\theta^2)=(\theta^1-c_5)^{c_4}(\theta^2-c_3)^2,\quad \theta^1> c_5,\theta^2\neq-c_3,-1<c_4<0.
\end{equation}
\end{itemize}
\item Let $\left(\frac{h'^2}{hh''}\right)'\neq0$.
\begin{itemize}
\item If $A_1(\theta^1)=0$, we have $A_3(\theta^1)=0$, namely $\frac{f'^2}{ff''}=1.$ The solution is
\begin{equation}
f(\theta^1)=c_3\exp(c_2\theta^1)
\end{equation}
and $h$ is arbitrary.
\item If $A_1(\theta^1)\neq0$, $\left(\frac{h'h'''}{h''^2}\right)'=c_1\left(\frac{h'^2}{hh''}\right)'$
and $A_3=-c_1A_1$ where $c_1\neq 0$. Simple calculation gives that
\begin{equation}
\frac{h'h'''}{h''^2}=c_1\frac{h'^2}{hh''}+c_2\label{case303}
\end{equation}
and
\begin{equation}
c_1\frac{f'f'''}{f''^2}=\frac{f'^2}{ff''}+c_1-1.\label{case302}
\end{equation}
These two equations are similar that solution of \eqref{case302} is simply a change of parameter in the solution of \eqref{case303}.  It is difficult to solve them but we can still find some special solutions. For instance, if we let $c_1=1,c_2=0$, both functions $f$ and $h$ are of the form 
\begin{equation}
c_4\exp(c_3\theta)+c_5\exp(-c_3\theta),
\end{equation}
where $\theta$ is $\theta^1$ or $\theta^2$ respectively.  For example, let us assume that all the constants are nonzero, then the function
\begin{equation}
\psi(\theta^1,\theta^2)=\left(c_1^2\exp(c_3\theta^1)+c_2^2\exp(-c_3\theta^1)\right)\left(c_4^2\exp(c_6\theta^2)+c_5^2\exp(-c_6\theta^2)\right)
\end{equation}
is convex.
\end{itemize}

\end{itemize}

When $\lambda\neq0$, the equation is even more complicated. It is still possible to obtain special solutions using the same algorithm above, however we do not provide details here.

In general, one may start with any ansatz and  solve the resulting equation \eqref{eeq}. For example, the ansatz $\psi(\theta^1,\theta^2)=f(\theta^1)h(\theta^2)+g(\theta^1)$ will lead to the potential function of the normal distribution in Example \ref{ex1c}. In principle, the solutions are  potential functions for statistical Einstein manifolds once the convexity condition is satisfied. 


\subsection{Group-invariant solutions}
In this part, we use the symmetry method to systematically consider group-invariant solutions of \eqref{eeq}. 
We rewrite the equation using a time-space convention by letting $\theta^1=t$ and $\theta^2=x$,
\begin{equation}\label{txpeq}
\psi_{tt}\left(\psi_{ttx}\psi_{xxx}-\psi_{txx}^2\right)-\psi_{tx}\left(\psi_{ttt}\psi_{xxx}-\psi_{ttx}\psi_{txx}\right)+\psi_{xx}\left(\psi_{ttt}\psi_{txx}-\psi_{ttx}^2\right)=4\lambda \left(\psi_{tt}\psi_{xx}-\psi_{tx}^2\right)^2.
\end{equation}
Note that the $x$ here is a space variable rather than the possible value of a random variable in a probability density function. The convexity condition is equivalent to the positive-definiteness of $g$, namely 
\begin{equation}
\psi_{tt}+\psi_{xx}> 0\text{ and } \det(g)=\psi_{tt}\psi_{xx}-\psi_{tx}^2> 0.
\end{equation}

Using the LSC condition \eqref{lsc}, we obtain a nine-dimensional Lie algebra of infinitesimal generators spanned by 
\begin{equation}
\begin{aligned}
&X_1=\partial_t,\quad X_2=\partial_x,\quad X_3=\partial_{\psi},\\
&X_4=t\partial_t,\quad X_5=x\partial_x,\quad X_6=x\partial_t,\\
&X_7=t\partial_x,\quad X_8=t\partial_{\psi},\quad X_9=x\partial_{\psi}.
\end{aligned}
\end{equation}
Next, we use (linear combinations of) them to calculate invariant solutions of potential functions for the equation \eqref{txpeq}.  The following list is not exhaustive; for instance, one may also consider a linear combination of multiple vector fields. Below we only present invariant solutions which are potential to be convex. Note that the constant $a$ is nonzero.

\begin{itemize}
 \item Using $X_4+aX_2$:
 
%
 
 \begin{equation}
 \psi(\theta^1,\theta^2)=-\frac {1 }{4\lambda}\ln  \left\{c_2\exp \left(c_1 \theta^2-c_1a\ln  \theta^1
  \right)-1\right\}+c_3;
 \end{equation}
 
  \item Using $X_5+aX_1$:
 
%
 
 \begin{equation}
 \psi(\theta^1,\theta^2)=-\frac {1 }{4\lambda}\ln  \left\{c_2\exp \left(c_1 \theta^1-c_1a\ln \theta^2
  \right)-1\right\}+c_3;
 \end{equation}
 
 \item Using $X_6+aX_2$:
 
%
 
 \begin{equation}
  \psi(\theta^1,\theta^2)=-\frac {1 }{4\lambda}\ln  \left\{c_2\exp \left(c_1 \left(\theta^2\right)^2-2c_1a  \theta^1
  \right)-1\right\}+c_3;
 \end{equation}
 
\item Using  $X_7+aX_1$:
 
%
 
 \begin{equation}
  \psi(\theta^1,\theta^2)=-\frac {1 }{4\lambda}\ln  \left\{c_2\exp \left(c_1 \left(\theta^1\right)^2-2c_1a  \theta^2
  \right)-1\right\}+c_3;
 \end{equation}
 
\item Using $X_8+aX_5$:

 \begin{equation}
 \psi(\theta^1,\theta^2)= \frac {    \left( \theta^1- c_1a\right)\ln \left( \theta^1-c_1a\right) - (1+4c_1\lambda) \theta^1\ln  \theta^1   }{4ac_1\lambda}+\frac{\theta^1\ln \theta^2  }{4a}+c_2\theta^1+c_3;
 \end{equation}

\item  Using $X_8+aX_6$: 

 \begin{equation}
 \psi(\theta^1,\theta^2)= \frac { \left(  \theta^2-c_1\right) \ln  \left( 
 \theta^2- c_1  \right) - \theta^2\ln \theta^2  }{4c_1 \lambda }+\frac{\left(\theta^1\right)^2}{2a\theta^2}  +
c_2 \theta^2+c_3;
 \end{equation}

%
 
 \item Using $X_9+aX_4$:
 
  \begin{equation}
 \psi(\theta^1,\theta^2)= \frac {    \left(\theta^2 - c_1a\right)\ln \left( \theta^2-c_1a\right) - (1+4c_1\lambda) \theta^2\ln  \theta^2   }{4ac_1\lambda}+\frac{\theta^2\ln  \theta^1  }{4a}+c_2\theta^2+c_3;
 \end{equation}
 
 \item Using $X_9+aX_7$:
  
   \begin{equation}
 \psi(\theta^1,\theta^2)= \frac { \left(  \theta^1-c_1\right) \ln  \left( 
 \theta^1- c_1  \right) - \theta^1\ln \theta^1  }{4c_1 \lambda }+\frac{\left(\theta^2\right)^2}{2a\theta^1}  +
c_2 \theta^1+c_3.
 \end{equation}
 
%

 
%


%

 \end{itemize}
 
 These invariant solutions are not necessary convex on the whole plane $\mathbb{R}^2$. It is always possible, however,  to construct convexity domains for the invariant solutions. In statistics, the parameters are also often defined on a restricted domain rather than the whole plane, e.g. the normal distribution in Example \ref{ex1c}.
 
\section{Conclusion}

In this paper, we have considered statistical Einstein manifolds of exponential families. Particularly in dimension two, this is equivalent to searching for constant curvature statistical manifolds. We used an ansatz method and symmetry method to construct potential functions of exponential families, whose constant curvature can be written using potential functions and their derivatives only. Nevertheless, there are challenging open questions to be investigated further.

One open question is to recover a probability density function from a given potential function, that we will call the inverse problem. As illustrated, we are always, at least in principle, able to get the potential function from a given pdf. It is not obvious if it is still possible and how we can do it vice versa. For instance, in dimension two, we will have to solve the following integral equation for unknown functions $F_1(x)$, $F_2(x)$ and $C(x)$,
\begin{equation*}
\exp\left(\psi(\theta^1,\theta^2)\right)=\int \exp\left\{F_1(x)\theta^1+F_2(x)\theta^2+C(x)\right\}\operatorname{d}\!x,
\end{equation*}
where the potential function $\psi(\theta^1,\theta^2)$ is given.
 
Another question is the investigation of higher dimensional statistical Einstein manifolds of exponential families, especially for dimensions greater than $4$ when the Einstein condition \eqref{eq} is not equivalent to constant-curvature condition any more.  Well-known higher dimensional exponential families include the multivariate normal distribution, the Dirichlet distribution, the Wishart distribution, the multinomial distribution and so on; see for instance, \cite{FEHP:2011,NG:2009}. \\

{\bf Acknowledgements.} LP is partially supported by JSPS Grant-in-Aid for Scientific Research (No. 16KT0024), the MEXT `Top Global University Project', Waseda University Grant for Special Research Projects (Nos. 2019C-179, 2019E-036, 2019R-081) and Waseda University Grant Program for Promotion of International Joint Research. ZZ is supported by Science and Technology Program of Beijing Municipal Commission of Education  (No. KM201810005006).

\end{document}